\documentclass[12pt,a4paper,english,superscriptaddress,aps,nofootinbib]{revtex4}
\usepackage[utf8]{inputenc}
\usepackage[T1]{fontenc}
\usepackage{amsmath,amssymb,graphicx}
\makeatletter
\usepackage{babel}
\usepackage[active]{srcltx}
\usepackage{graphicx,color}
\usepackage{changebar}
\usepackage{hyperref}
\usepackage[T1]{fontenc}
\usepackage{esint}
\usepackage{multirow}
\usepackage{dsfont}
\usepackage{ae}
\usepackage{amsmath}
\usepackage{braket}
\usepackage{mathtools}
\usepackage{slashed}
\usepackage{empheq}
\usepackage{multirow}
\usepackage{graphicx}
\usepackage{amsfonts}
\usepackage{amsmath}
\newcommand{\e}{\text{e}}
\newcommand{\be}{\begin{equation}}
	\newcommand{\ee}{\end{equation}}
\newcommand{\bea}{\begin{eqnarray}}
	\newcommand{\eea}{\end{eqnarray}}

\begin{document}

\title{Aspects of kink-like structures in 2D dilaton gravity}

\author{F. C. E. Lima}
\email{cleiton.estevao@fisica.ufc.br}
\affiliation{Programa de P\'{o}s-gradua\c{c}\~{a}o em F\'{i}sica, Universidade Federal do Maranh\~{a}o, Campus Universit\'{a}rio do Bacanga, S\~{a}o Lu\'{i}s, Maranh\~{a}o, 65080-805, Brazil.}
\affiliation{Departamento do F\'{i}sica, Universidade Federal do Cear\'{a}, Campus do Pici, Fortaleza, Cear\'{a},  60455-760, Brazil.}

\author{C. A. S. Almeida}
\email{carlos@fisica.ufc.br}
\affiliation{Departamento do F\'{i}sica, Universidade Federal do Cear\'{a}, Campus do Pici, Fortaleza, Cear\'{a}, 60455-760, Brazil.}
\affiliation{Institute of Cosmology, Department of Physics and Astronomy, Tufts University, Medford, Massachusetts, 02155, USA.}

\begin{abstract}\vspace{0.4cm}
\noindent \textbf{Abstract:} The topological structures that arise from two-dimensional models are relevant physically and the first step towards understanding more complex systems. In this work, one studies the kink-like solutions of the matter field that emerge in a two-dimensional dilaton gravity scenario. Considering this scenario, we examine the linear stability of the matter field and the translational mode. Due to the specific profile of the solutions found, a differential configurational complexity (DCC) analysis of the system was necessary to confirm the nature of the structures. Furthermore, we study the interforce and scattering process, considering a pair of kink-antikink-like solutions in a fixed time ($t=0$).

\end{abstract}
\maketitle

\thispagestyle{empty}


\section{Introduction} 

It is notorious that high-dimensional cosmological models suggest some explanations for open-ended problems in high-energy physics\cite{Bronnikov}.  These models can provide us with interpretations of open-ended issues in physics as the problem related to the hierarchy \cite{LS,Arkani}, the origin of dark matter \cite{Arkani1}, the cosmological constant \cite{Chen}, and cosmic acceleration \cite{Khoury,sr}. However, a simple alternative to understanding these cosmological models is to analyze the two-dimensional (2D) gravitational models \cite{Zhong}. In the last decades, the study of two-dimensional models has shown promise and helped understand the effects of quantum gravity \cite{Henneaux,Alwis}, gravitational collapse \cite{Vaz,Vaz1}, and black hole evaporation \cite{Callan,Bilal,Russo}. More recently, the study of self-gravitating kinks in 2D dilaton gravity \cite{Zhong} has contributed to the return of the interest on these structures (for more applications, see Refs. \cite{Grumiller,Ishii,Hartman}).

An immediate way to extend Einstein's gravity in 2D is to introduce a dilaton field with non-minimal coupling with the metric. Immediately, Teitelboim \cite{Teitelboim} and Jackiw \cite{Jackiw} made an extension of Einstein's theory into 2D models. Soon after, an extension of the Jackiw-Teitelboim action, Mann, Morsink, Sikkema, and Steele (MMSS) proposed an approach in which the contribution of the cosmological constant is absent \cite{Mann}. In the MMSS model, the dilaton field has a dynamic term that leads to energy-momentum conservation. The MMSS action is
\begin{align}
    S_{\text{MMSS}}=\frac{1}{\kappa}\int\, d^2x\, \sqrt{-g}\bigg[-\frac{1}{2}(\nabla\phi)^2+\phi R+\kappa\mathcal{L}_{M}\bigg].
\end{align}
The MMSS action is interpreted as a limit when $D\to 2$ for general relativity. The fact is that models governed by an MMSS-like action have been shown effective for studies of black hole chemistry and entropic gravity. On the other hand, the study of the dynamics of the matter field is still a little-frequented area. Motivated by these discussions, we will, throughout this work, study the dynamics of self-gravitating topological structures in two-dimensional dilaton gravity.

The action of MMSS is interpreted as a limit when $D\to 2$ for general relativity. The fact is that models governed by an MMSS type action have been shown effective for studies of black hole chemistry \cite{Frassino} and entropic gravity \cite{Mann1}. On the other hand, the study of the dynamics of the matter field is still a little-frequented area. Motivated by these discussions, we will, throughout this work, study the dynamics of the self-gravitating topological structures in two-dimensional dilaton gravity.

The main feature that motivates the studies of 2D structures is that, in low dimensions, we obtain the field by elementary techniques. Moreover, one can describe several physical systems approximately or effectively assuming one-dimensional field configurations \cite{Vakhid}. Additionally, to reach the 2D field configurations, the presence of domain walls in the theory is necessary. On the other hand, to confirm the existence of the domain wall, we can use arguments emerging from the information theory, i. e., the configurational entropy (CE) and its variants. Several works, in various dimensions, have used fundamentals from the CE to predict and complement the understanding of topological structures \cite{RRocha,RRocha1}. This interpretation is possible since these formalisms can provide criteria to control the stability of field configurations based only on the informative content. For a review of the relevant features and applications of the CE and their properties to predict possible phase transitions, see the papers of Gleiser et al. \cite{Gle,Gle1,Gle2,Gle3}.

Recently, have been performed several studies on 2D structures \cite{Manton2,Zlo1,Zlo2,Sugiyama,Campbell}. This interest is due to the diversified results that arise in the inquiry of the kink structures. Some impressive and promising results about 2D configurations appear in studies of the long-range interactions between kink and antikink \cite{Vakhid1,Belen1,Manton1}. Other attractive results involving these topological structures also appear in the study of point particles and zero-branes \cite{Zlo3,Zlo4}, in domain walls \cite{Vachaspati}, and investigations on the emergence of the universe \cite{VAGani}. Thus, one can conclude that these researches on 2D configurations are of great interest and help us understand more complex problems in theoretical physics \cite{Moreira,Ranieri1}.

Our central purpose in this paper is the study of the emergence of self-gravitating topological structures in two-dimensional dilaton gravity. Then, one seeks to understand the influence of the dilaton field and the spacetime curvature on the matter field. Furthermore, we investigate which internal force, or long-range interforce, causes this interaction. The scattering process produced by this interforce also is examined. Finally, adopting arguments from CE, the phase transitions and the class of solutions of the structure are investigated.

We organize our work as follows: We discuss the model and topological solutions of the matter field and its stability in section II. In section III, one calculates the inter-kink force and investigates the structure scattering. Posteriorly, we study the phase transitions of the matter field in section IV. Finally, in section V, our findings are announced.

\section{The model and its stability} 

We know from the literature that the Einstein tensor in 2D is null. Thus, to describe a model in 2D gravity, it is necessary to extend the Einstein-Hilbert action. An immediate way to extend Einstein's gravity occurs when introducing a dilaton field coupled non-minimally with the metric. Given this consideration, let us consider the MMSS-type action \cite{Mann}:
\begin{align}
    \label{eq1}
S=\frac{1}{\kappa}\int\, d^2 x\, \sqrt{-g}\bigg[-\frac{1}{2}\nabla_\mu\phi\nabla^\mu\phi+\phi R+\kappa\bigg(-\frac{1}{2}\nabla_\mu\psi\nabla^\mu\psi-V(\psi)\bigg)\bigg],    
\end{align}
where $R$ is the Ricci scalar, $\kappa$ is the gravitational coupling, $\phi$ is the dilaton field, and $\psi$ is a real scalar field.

The variation of action leads to equations of motion, i. e., the Einstein equation
\begin{align}\label{0}
    \kappa T_{\mu\nu}= -\nabla_{\mu}\phi\nabla_\nu\phi+\frac{1}{2}g_{\mu\nu}(\nabla^{\rho}\phi\nabla_{\rho}\phi+4\nabla_{\rho}\nabla^{\rho}\phi)-2\nabla_{\mu}\nabla_{\nu}\phi, 
\end{align}
the dilaton field equation
\begin{align}\label{1}
    \nabla^{\mu}\nabla_{\mu}\phi+R=0, 
\end{align}
and the scalar field equation
\begin{align}\label{2}
    \nabla^{\mu}\nabla_{\mu}\psi+\frac{dV}{d\psi}=0. 
\end{align}

The stress-energy tensor is 
\begin{align}
    \label{tem}
    T_{\mu\nu}=\nabla_{\mu}\psi\nabla_{\nu}\psi-\frac{1}{2}g_{\mu\nu}(\nabla^{\rho}\psi\nabla_{\rho}\psi+2V).
\end{align}

For our study of the two-dimensional structures, let us consider the metric given by
\begin{align}
    \label{eq2}
    ds^2=\text{e}^{2A(r)}(-dt^2+dr^2).
\end{align}
The metric (\ref{eq2}) is widely used in the study of topological structures in extra dimensions \cite{Moreira,Yang}. The term $A(r)$ is the warp function.

Substituting the metric (\ref{eq2}) in the static Einstein equation (\ref{0}), one obtains
\begin{align}\label{e1}
    2\phi''(r)+\phi'(r)^2-2A'(r)\phi'(r)=\kappa \psi'(r)^2,
\end{align}
and
\begin{align}\label{e2}
    \phi''(r)+A'(r)\phi'(r)=-\kappa V(\psi). 
\end{align}
Analyzing the Eqs. (\ref{1}), (\ref{2}), (\ref{e1}), and (\ref{e2}), one can conclude that only three of them are linearly independent. Indeed, we can derive the Eq. (\ref{e2}) using the Eqs. (\ref{1}), (\ref{2}) and (\ref{e1}). Besides, one way to solve the above equation system is to use the first-order formalism, see Ref. \cite{MLSC, CGLM}. In first-order formalism, it assumes a superpotential related to the interaction of the matter field to reduce the order of the equations of motion. In this work, we will not use first-order formalism, i. e. we will solve the second-order equations of the system numerically.

For simplicity, it is convenient to write the action (\ref{eq1}) in terms of the fields $\phi(r)$, $\psi(r)$, and $A(r)$. For this, we note that the Ricci scalar is
\begin{align}
    \label{eq3}
    R=-2\text{e}^{-2A(r)}A''(r).
\end{align}
Therefore, the action is
\begin{align}
    \label{eq4}
    S=&\frac{1}{\kappa}\int\, d^2x\, \bigg[-\frac{1}{2}\phi{{'}(r)}^{2}-2A''(r)\phi(r)-\frac{1}{2}\kappa\psi^{'}(r)^{2}-\kappa\text{e}^{2A(r)}V(\psi)\bigg],
\end{align}
where the prime notation is the derivative regarding the position variable.

For the topological structures to arise is necessary to preserve the spontaneous symmetry breaking. Thus, we adopt a $\varphi^4$-like interaction, i.e.,
\begin{equation}\label{ewq}
    V(\psi)=\frac{\lambda}{2}(v^2-\psi^2)^2.
\end{equation}

Allow us to start from the assumption that we know the warp function. Consequently, action happens to have two fields with the independent variable $r$. These fields will describe the matter and dilaton fields. The equations of motion of scalar and dilaton fields are, respectively, given by
\begin{align}
    \label{eq5}
    \psi''(r)+2\lambda\text{e}^{2A(r)}\psi(r)(v^2-\psi(r)^2)=0,
\end{align}
and
\begin{align}
    \label{eq6}
    \phi''(r)-2A''(r)=0.
\end{align}

From Eq. (\ref{eq6}), we obtain a relation between the dilaton field and the warp function, namely,
\begin{align}\label{warp}
    \phi(r)=2A(r)+\beta r+\gamma.
\end{align}
To preserve the translational invariance of the $\psi(r)$ field is convenient to assume the condition $\beta=\gamma=0$. This condition is appropriate because it also allows describing a topological matter field\footnote{Topological matter fields are fields in which the topological boundary conditions are valid. In this case, these conditions are $\psi(r\to\pm\infty)\to\pm v$.
}. For more details, see Refs. \cite{Zhong,Vachaspati}.

Moreover, due to the choice of potential in Eq. (\ref{ewq}), the warp factor assumes the form
\begin{align}
    \label{eq8}
    \text{e}^{2A(r)}=\cosh^{-2p}(\lambda_0 r), 
\end{align}
where the $\lambda_0$-parameter makes the argument of hyperbolic dimensionless. Indeed, one can assume $\lambda_0=1$. In 5D braneworlds, this profile of the warp function is widely used \cite{Moreira,Ranieri1}. In Refs. \cite{Moreira,Ranieri1}, using this profile of the warp function, the authors demonstrated the symmetrical and asymmetrical solutions for a thick brane in 5D. In our 2D case, the warp factor (\ref{eq8}) leads us to the following equation for the kink structures
\begin{align}\label{eq9}
    &\psi''(r)+2\lambda\cosh^{-2p}(r)\psi(r)(\nu^2-\psi(r)^2)=0. 
\end{align}

\begin{figure}[ht!]
    \centering
    \includegraphics[height=7cm,width=8cm]{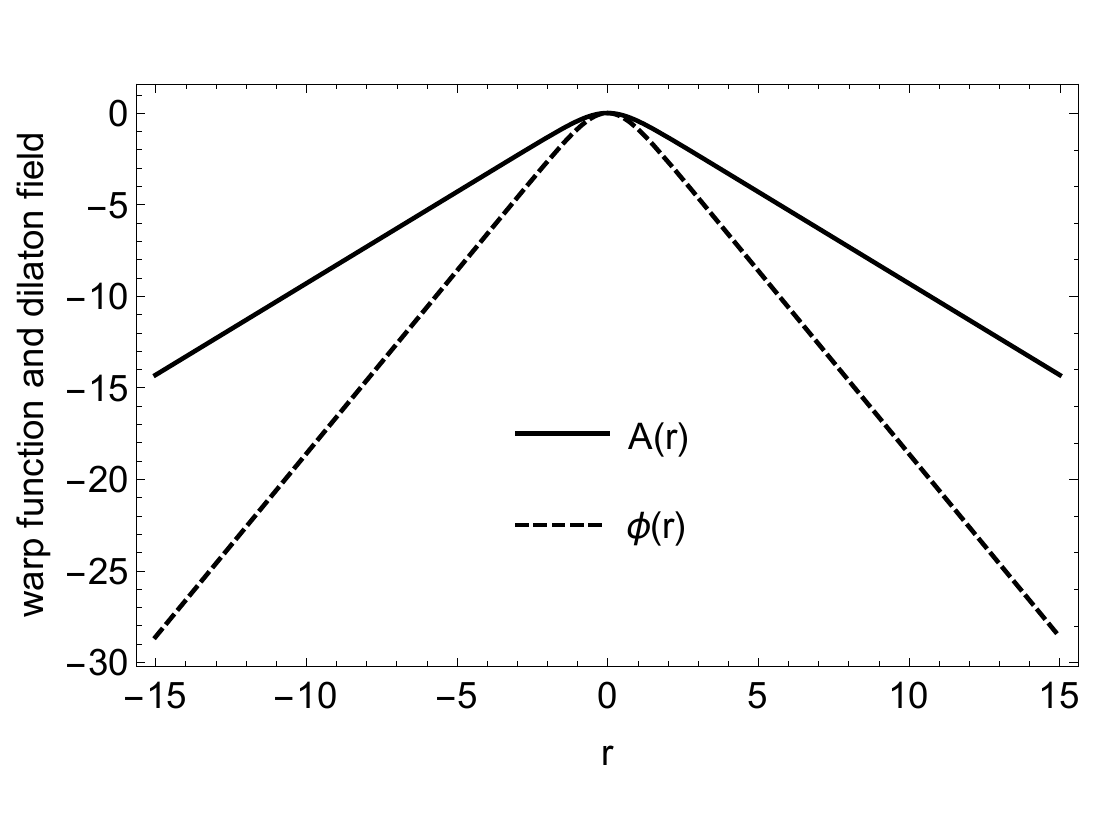}\vspace{-0.3cm}
    \caption{Warp function and the dilaton field.}
    \label{fig1}
\end{figure}

For the metric (\ref{eq2}) and the warp factor (\ref{eq8}), the dilaton field profile is given by
\begin{align}
    &\phi(r)=\text{ln}(\cosh^{-2p}(r)).
\end{align}
We displayed the profile of the dilaton field in Fig. \ref{fig1}. Note that our results for the dilaton field behavior are similar to those found in Ref. \cite{Stoetzel}.

Eq. (\ref{eq9}) is easily solved using numerical techniques. For simulation, we use topological boundary conditions, i. e.
\begin{align}\label{eq11}
    \psi(r\to \pm\infty)=\pm \nu.
\end{align}
where, for convenience, one adopts $\nu=1$.

\begin{figure}[ht!]
    \centering
    \includegraphics[height=7cm,width=8cm]{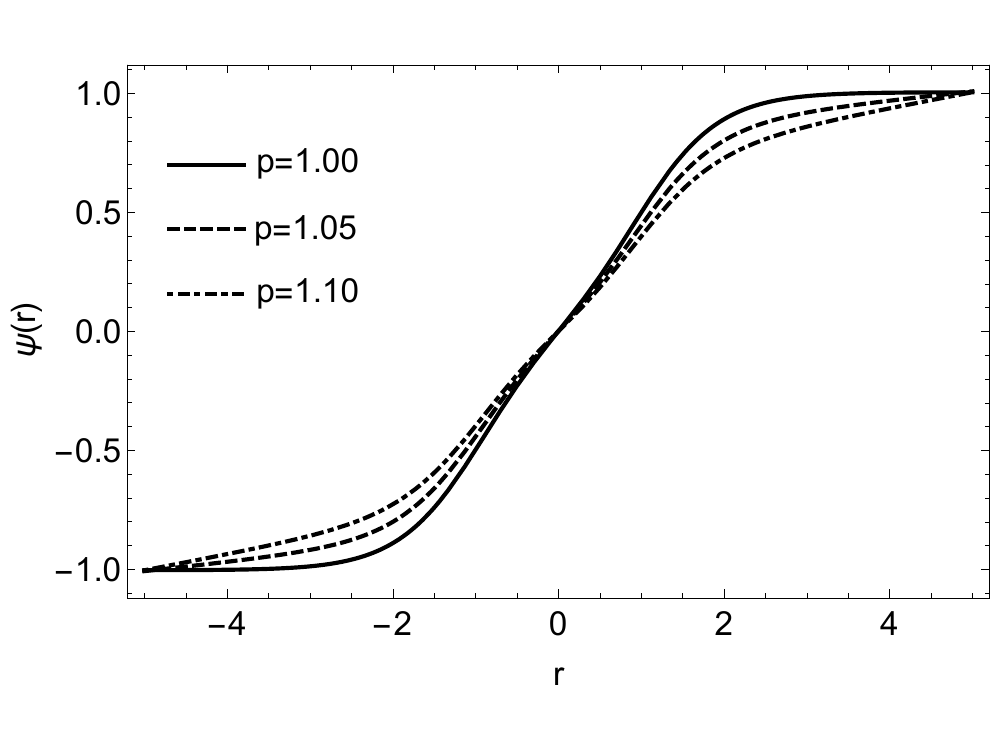}
    \includegraphics[height=7cm,width=8cm]{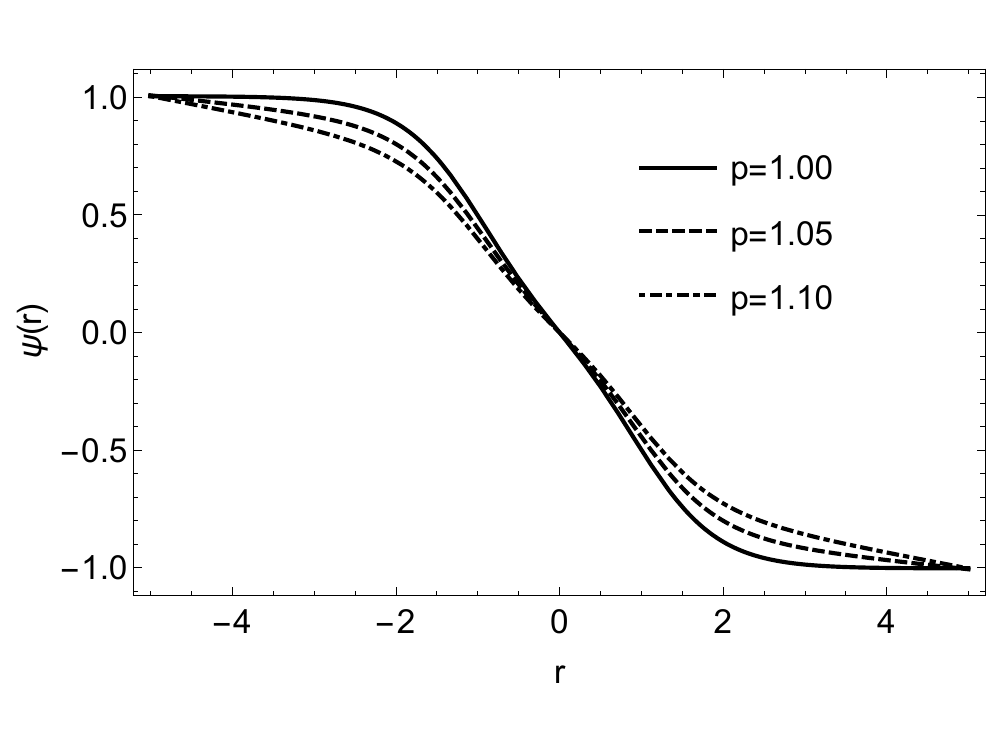}\\ \vspace{-0.3cm}
        \begin{center} \; \; (a) \hspace{7.5cm} (b)\end{center}\vspace{-0.3cm}
    \caption{Matter field when $p=1.00$, $1.05$, and $1.15$.}
    \label{fig2}
\end{figure}

As a result, the configuration of the y field found has a double-kink-like profile. We expose the double-kink-like solutions in Fig. \ref{fig2}(a). Meanwhile, the double-antikink-like configurations are displayed in Fig. \ref{fig2}(b).

As a result, the configuration of the $\psi(r)$ field found has a double-kink-like profile. We expose the double-kink-like solutions in Fig. \ref{fig2}(a). Meanwhile, the double-antikink-like configurations are displayed in Fig. \ref{fig2}(b).

\subsection{The stability of the solutions} 

Let us analyze the linear stability of the kink-antikink-like self-gravitating solutions performing small perturbations in the field solutions (Fig. \ref{fig2}). Allow us to consider this small perturbation around arbitrary static solutions. This perturbation of the metric is given by
\begin{align}
    \delta g_{\mu\nu}=\text{e}^{2A(r)}\begin{pmatrix}
    h_{00}(r,t) & \Phi(r,t)\\
    \Phi(r,t) & h_{rr}(r,t)
    \end{pmatrix},
\end{align}
and for the first order, the perturbation of the inverse metric is
\begin{align}
    \delta g^{\mu\nu}=-\text{e}^{-2A(r)}h^{\mu\nu}.
\end{align}
As assumed in Refs. \cite{Zhong,Zhong1}, it is convenient to consider a new variable, $\Xi=2\dot{\Phi}-h^{'}_{00}$, and the dilaton gauge as $\delta\phi= 0$.

The linearization of Einstein's equation leads to
\begin{align}
&2A'\delta\phi'-2A'\phi'h_{rr}-2\delta\phi''-\delta\phi'\phi'+h_{rr}^{'}\phi^{'}+2h_{rr}\phi''+\frac{1}{2}h_{rr}\phi^{_{'}\,^{2}}=\kappa\bigg(\psi'\delta\psi'+\psi''\delta\psi-\frac{1}{2}\psi^{_{'}\,^{2}}h_{rr}\bigg),\\
& 2A'\delta\phi-2\delta\phi'-\phi'\delta\phi+\phi'h_{rr}=\kappa\psi'\delta\psi,\\
& 2A'\delta\phi'-2A'\phi'h_{rr}-\delta\phi'\phi'-\Ddot{\delta}\phi+\frac{1}{2}h_{rr}\phi^{_{'}\,^{2}}+\Xi\phi'=\kappa\bigg(\psi'\delta\psi'-\psi''\delta\psi-\frac{1}{2}\psi^{_{'}\,^{2}}h_{rr}\bigg).
\end{align}

From the linearization of the scalar field (\ref{2}), we found a new perturbation equation of the matter field, namely, 
\begin{align}
    \Ddot{\delta}\psi+\delta\psi''+2A'\frac{\psi''}{\psi}\delta\psi-\frac{\psi'''}{\psi'}\delta\psi-\frac{1}{2}\psi'h^{'}_{rr}-\psi''h_{rr}+\frac{1}{2}\psi'\Xi=0.
\end{align}

Note that we have three independent perturbation equations. Nonetheless, we should notice that the perturbation variables are not all independent. By analyzing the invariance of the perturbed equations, it comes to
\begin{align}\label{Schro}
    \Ddot{\delta}\psi-\delta\psi''+V_{_{\text{eff}}}(r)\delta\psi=0.
\end{align}
For more details on perturbative calculations, see Refs. \cite{Zhong,Zhong1}. 

\begin{figure}[ht!]
    \centering
    \includegraphics[height=7.5cm,width=8cm]{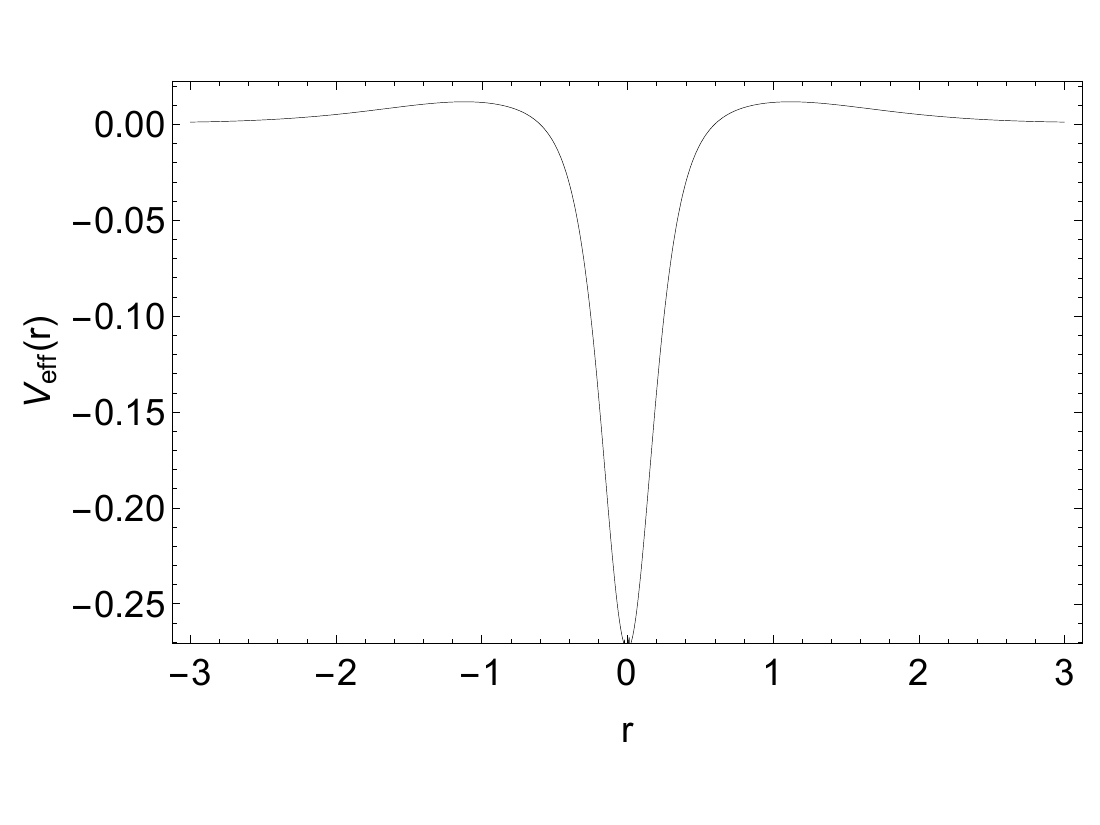}
    \includegraphics[height=7cm,width=8cm]{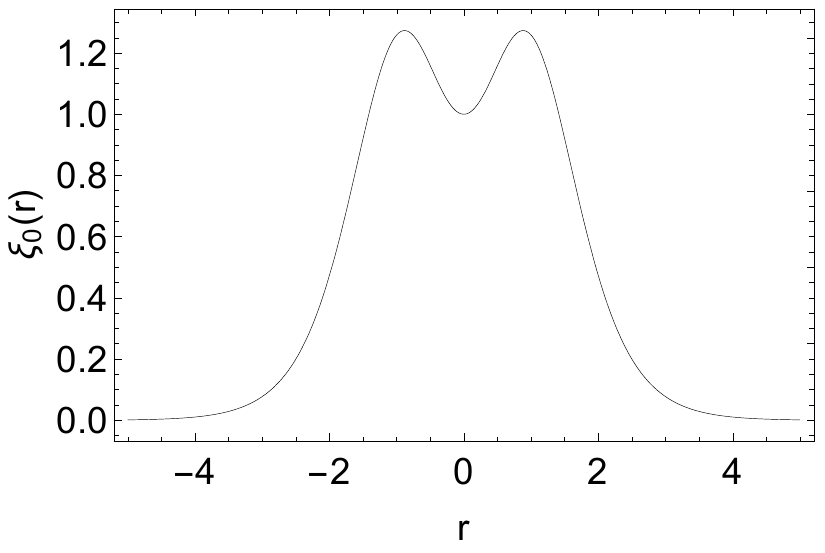}\\ \vspace{-0.3cm}
    \begin{center} \; \; (a) \hspace{7.5cm} (b)\end{center}\vspace{-0.3cm}
    \caption{(a) Effective potential. (b) Translational mode.}
    \label{fig3}
\end{figure}

The effective potential is given by
\begin{align}\label{pot}
    V_{_{\text{eff}}}(r)=4A''-2A'\frac{\psi''}{\psi'}-\phi''+2\bigg(\frac{\phi''}{\phi'}\bigg)^{2}-2\frac{\phi'''}{\phi'}+\frac{\psi'''}{\psi'}.
\end{align}
This effective potential is illustrated in Fig. \ref{fig3}(a).

Let us now, as usual, assume that $\delta\psi=\xi(r)$e$^{i\omega t}$. This consideration leads us to
\begin{align}\label{Sch}
    -\xi''(r)+V_{_{\text{eff}}}(r)\xi(r)=\omega^2\xi(r).
\end{align}

By arguments similar to those of supersymmetric quantum mechanics (see Refs. \cite{Zhong,Vakhid1,Zhong1}), one obtains that the translational mode of the kink-like structure is
\begin{align}
    \xi_0(r)=\mathcal{C}_0\frac{\psi'(r)}{2A'(r)},
\end{align}
where $\mathcal{C}_0$ is a normalization constant. The translational mode of the kink-like structure was obtained and exposed in figure \ref{fig3}(b).

It is interesting to highlight some similarities of our model with results of works in extra-dimensions. For example, in braneworlds, the effective potentials have singularities. These singularities suggest that the scalar modes are non-normalizable. Even as models in 5D, our theory appears to be free of long-range scalar force problems and indicates the nonexistence of resonance. For details on this discussion, please see Refs. \cite{Yang,Zhong3,Zhong4}. We will confirm this scattering result in Sec. IV. A peculiar feature arises in the study of the translational mode. In truth, the zero modes behave as if the topological structure were described by a double-kink-like configuration, thus admitting multiple-phase transitions. We will explore this result when discussing the DCC of the matter field. 

\section{Inter-kink force} 

In this section, we are interested in studying the force which acts between structure pairs with opposite topological charge. Our interest in the interforce of this system is because this investigation can give us information about the process of structure scattering\footnote{We will study the scattering of this structure Sec. 3.1.} \cite{Vachaspati}.

The inter-kink force or interforce is an internal force between topological structures as they approach each other. Thus, this force can induce a scattering process (or not) between the topological structures. If the interforce between configurations is attractive, a scattering process will occur in the dynamic case \cite{Vachaspati}. Particularly, the scattering phenomenon plays a relevant role in topological field theories because, from this process, it is possible to state whether a structure is a true soliton \cite{Vachaspati}.

\begin{figure}[ht!]
    \centering
    \includegraphics[height=7cm,width=8cm]{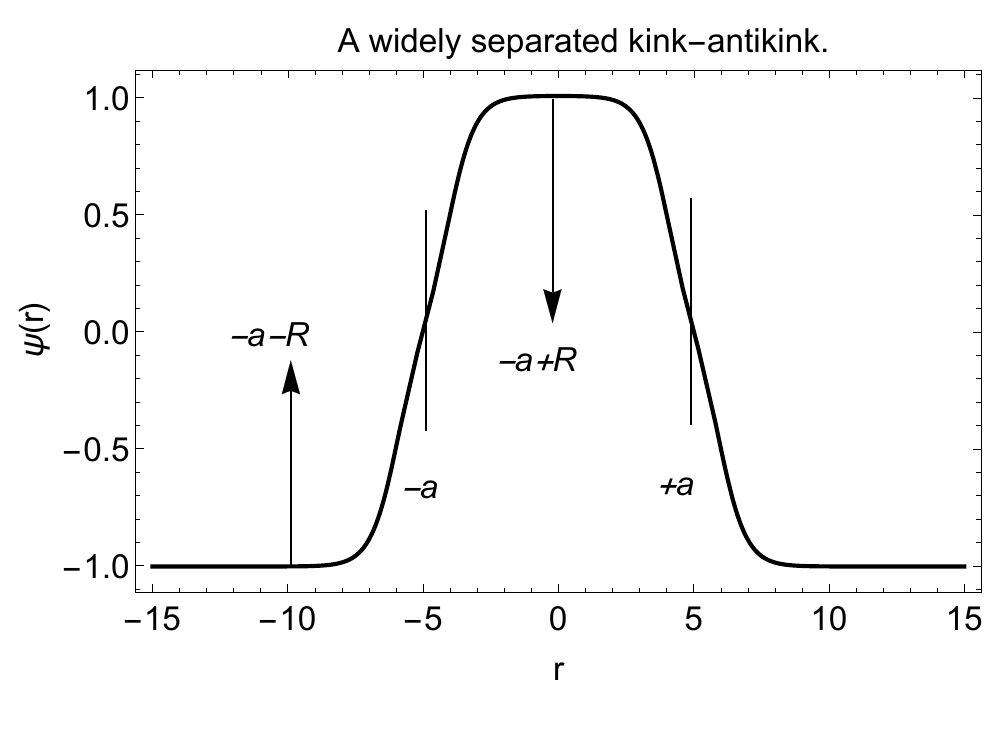}   \vspace{-0.3cm}
    \caption{Interaction of the Kink-antikink-like widely separated in fixed time.}
    \label{fig4}
\end{figure}

To study the interforce, let us start by assuming that our solutions (Fig. \ref{fig2}(a) and Fig.  \ref{fig2}(b)) can form a pair of kink-antikink-like structures widely separated. The spacing between these structures is $2a$. For convenience, the combination of the solutions of Eq. (\ref{eq9}) is exhibited in Fig. (\ref{fig4}).

The force exerted on a topological structure must obey Newton's second law. To calculate this force, we must analyze the momentum of the field. The integral of the momentum density $T^{0i}=\text{e}^{4A(r)}T_{0i}$ will give us the momentum of the structure in a large region around the kink. For the configuration located at $x=-a$, we will choose to look at the localized field momentum at $(-a-R,-a+R)$, i. e.,
\begin{align}
    P=\int_{-a-R}^{-a+R}\, dr\, \cosh^{-4p}(r)\dot{\psi}(r,t)\psi'(r,t),
\end{align}
where the dot notation denotes the time derivative.

The force on the field in this region is given by
\begin{align}\label{f}
F=\frac{dP}{dt}=\int^{-a+R}_{-a-R}\, \cosh^{-4p}(r) [\Ddot{\psi}(r,t)\psi'(r,t)+\dot{\psi}(r,t)\dot{\psi}'(r,t)]\, dr.
\end{align}

Remembering that
\begin{align}\label{f1}
    \Ddot{\psi}(r,t)=\psi''(r,t)-2\lambda\cosh^{-2p}(r)\psi(r,t)(\nu^2-\psi(r,t)^2),
\end{align}
and by replacing (\ref{f1}) in (\ref{f}), one obtains
\begin{align}
    \label{f2} 
        F=\int^{-a+R}_{-a-R}\, \cosh^{-4p}(r)\, \bigg[\frac{1}{2}\frac{d\psi'(r,t)^2}{dr}+\cosh^{-2p}(r)\frac{dV}{dr}+\dot{\psi}(r,t)\dot{\psi}'(r,t)\bigg]\, dr.
\end{align}

To finish our considerations, we will assume that the configuration in figure (\ref{fig4}) is initially static. This consideration will tell us that
\begin{align}
    \partial_t\psi\vert_{t=0}=0,
\end{align}
what gives us
\begin{align}\label{f4}
    F=\int^{-a+R}_{-a-R}\, \cosh^{-4p}(r)\, \bigg[\frac{1}{2}\frac{d\psi'(r,t)^2}{dr}+\cosh^{-2p}(r)\frac{dV}{dr}\bigg]\, dr,
\end{align}
where $V$ is the interaction.

We can numerically calculate the structure interforce. To reach our purpose, i.e., for simulation issues, we will consider $\vert a\vert=5$ and $\vert R\vert=4$. To calculate the force, we substitute the numerical solutions of the matter field $\psi(r)$ (solutions shown in Fig. \ref{fig2}) in Eq. (\ref{f4}).

The numerical result of the magnitude of the interforce generated by the kink-antikink-like interaction shown in Fig. \ref{fig4}, is displayed in the table (\ref{tab1}) for various values of the $p$-parameter.

\begin{table}[ht!]
\centering
\caption{Numerical results of the interforce between kink-antikink-like structures.}
\label{tab1}
\resizebox{6cm}{2.5cm}{%
\begin{tabular}{|c|c|}\hline
$p-$parameter & interforce  \\ \hline
  1.00 & $-0.1023\cdot 10^0$ \\ \hline
  1.05 & $-3.6197\cdot 10^{-9}$ \\ \hline
  1.10 & $-3.2558\cdot 10^{-11}$ \\ \hline
\end{tabular}}
\end{table}

\begin{figure}[ht!]
    \centering
    \includegraphics[height=7cm,width=8cm]{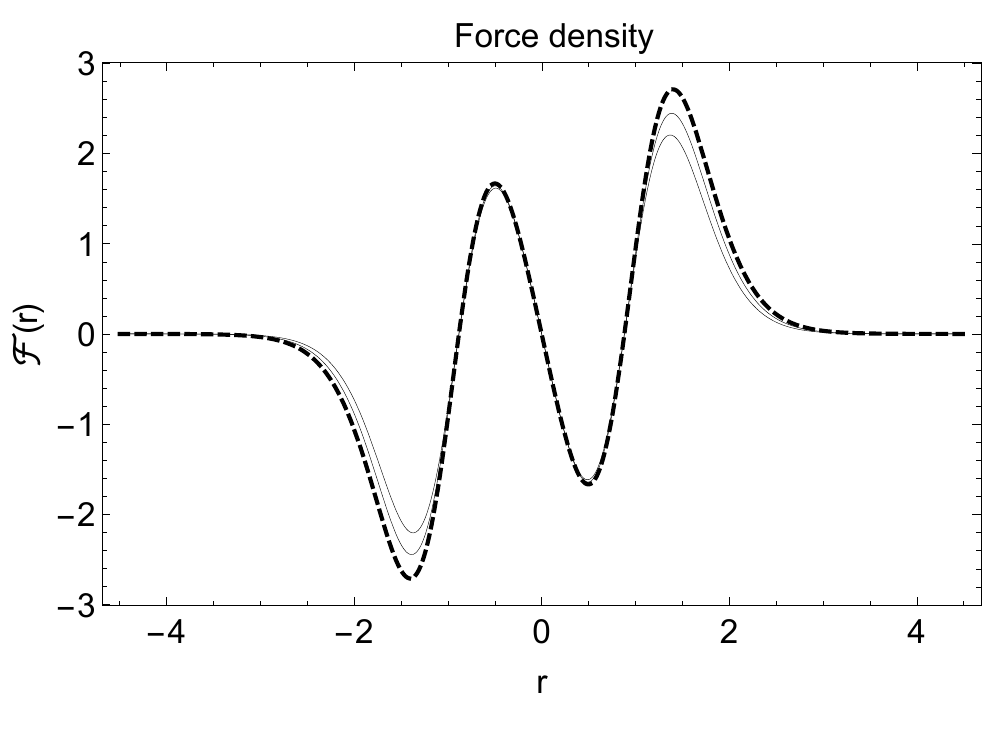}  \vspace{-0.75cm}
    \caption{Interforce density exerted on topological structures.}
    \label{fig5}
\end{figure}

The plot of the interforce density (\ref{f4}) is shown in Fig. (\ref{fig5}). Note that in the boundary region, where kink-antikink-like structures interact, the force density profile becomes ``oscillating'' with a sharp change in the amplitude of the interforce density. Note that the interforce density in the center of the structures (i. e., at $r=0$) is null. In addition, note that the smaller the p-parameter, the smaller the contribution of the warp factor, and the smaller the interforce intensity.

\subsection{Structure scattering phenomenon}

\begin{figure}[ht!]
    \centering
    \includegraphics[height=7cm,width=8cm]{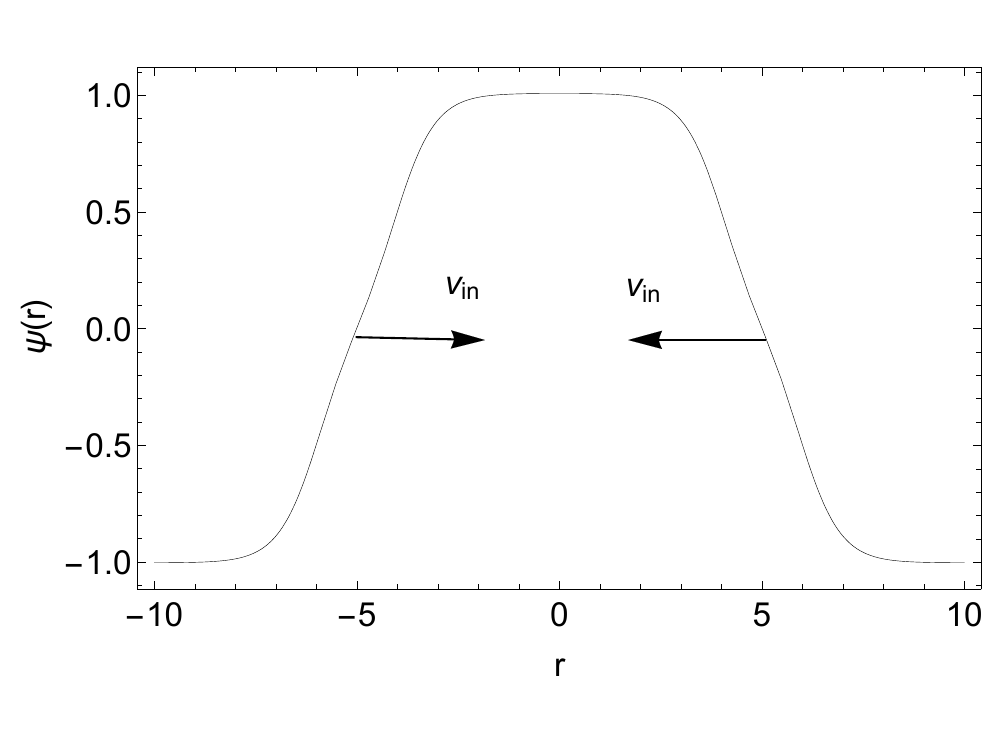}  \vspace{-0.7cm}
    \caption{Representation of the field configurations before the scattering.}
    \label{SGfig6}
\end{figure}

Allow us to study the scattering process of the kink-antikink solutions that our model support. In order, we investigate the equations of motion considering the dynamic case. In this scenario, one finds two solutions again, i. e., kink-like and antikink-like configurations. Numerically, we assume that the kink structures are initially at $r=+a$. Meanwhile, the antikink-like solutions are at $r=-a$. Furthermore, one considers, initially, that these structures move in opposite directions with initial velocity $v_{\text{in}}$. Doing this procedure, we study the scattering of structures. We present the process result in figure \ref{SGfig6}.

\begin{figure}[ht!]
    \centering
    \includegraphics[height=6cm,width=7.5cm]{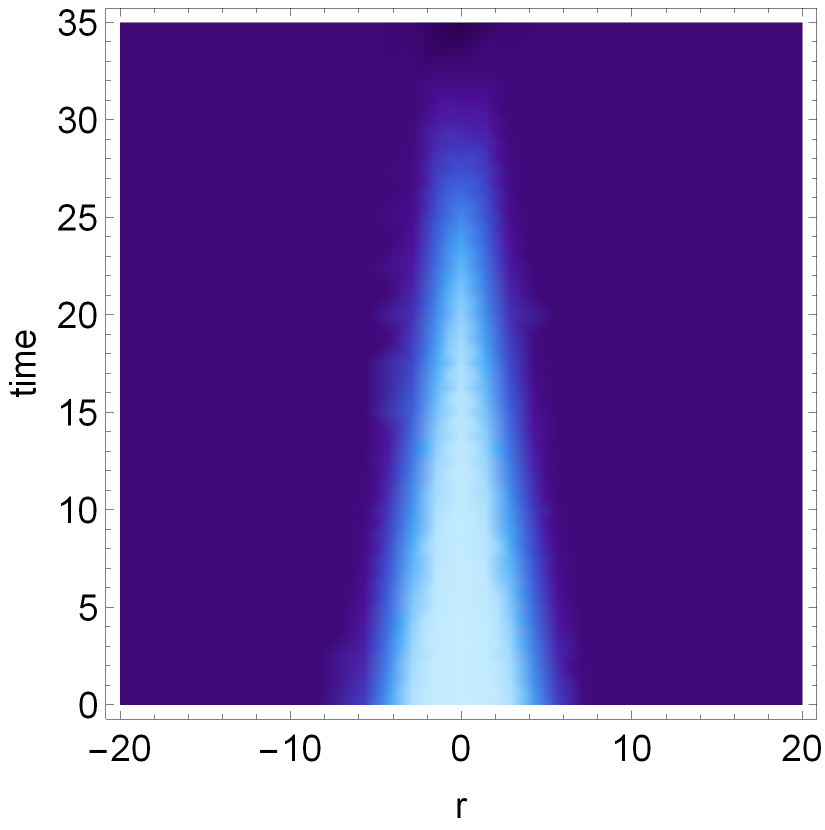}
    \includegraphics[height=6cm,width=7.5cm]{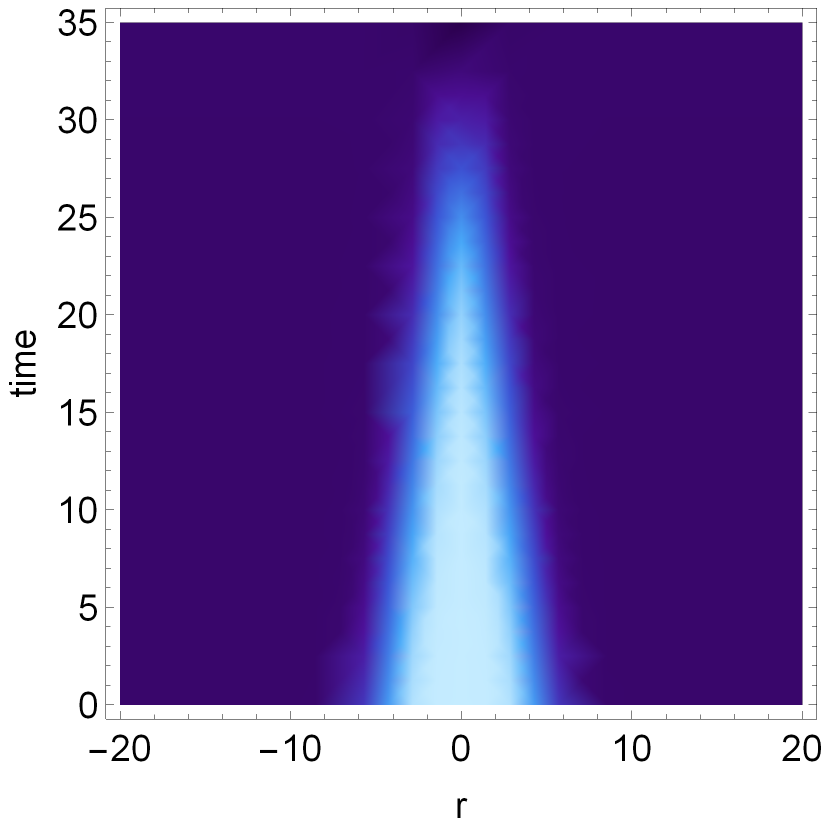}\\
       (a) $v_{in}=0.15$ \hspace{5cm} (b) $v_{in}=0.30$\\
    \includegraphics[height=6cm,width=7.5cm]{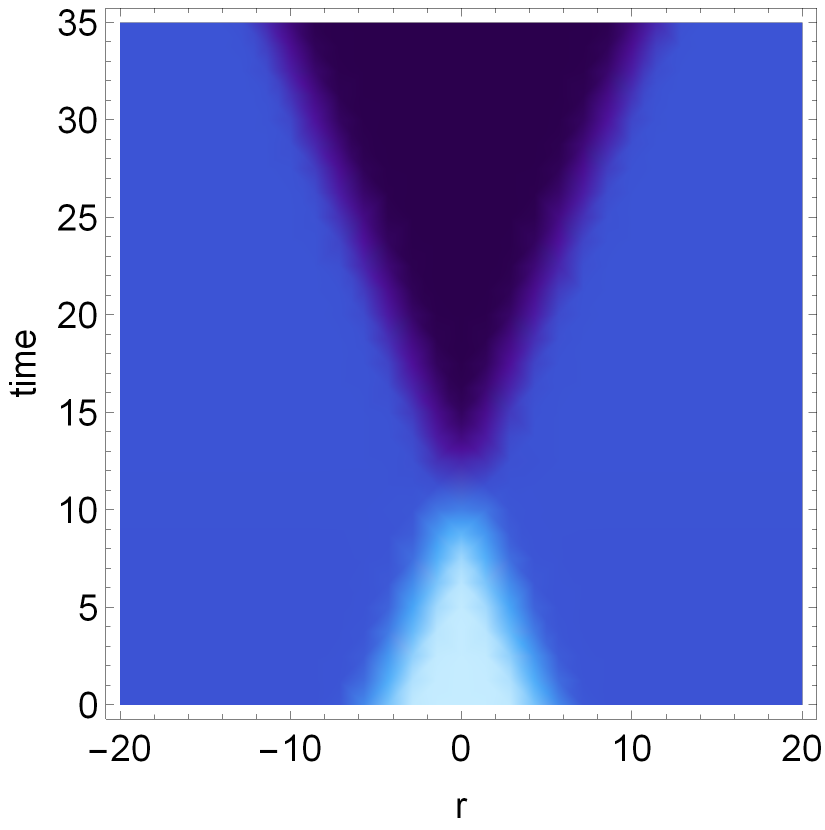}
    \includegraphics[height=6cm,width=7.5cm]{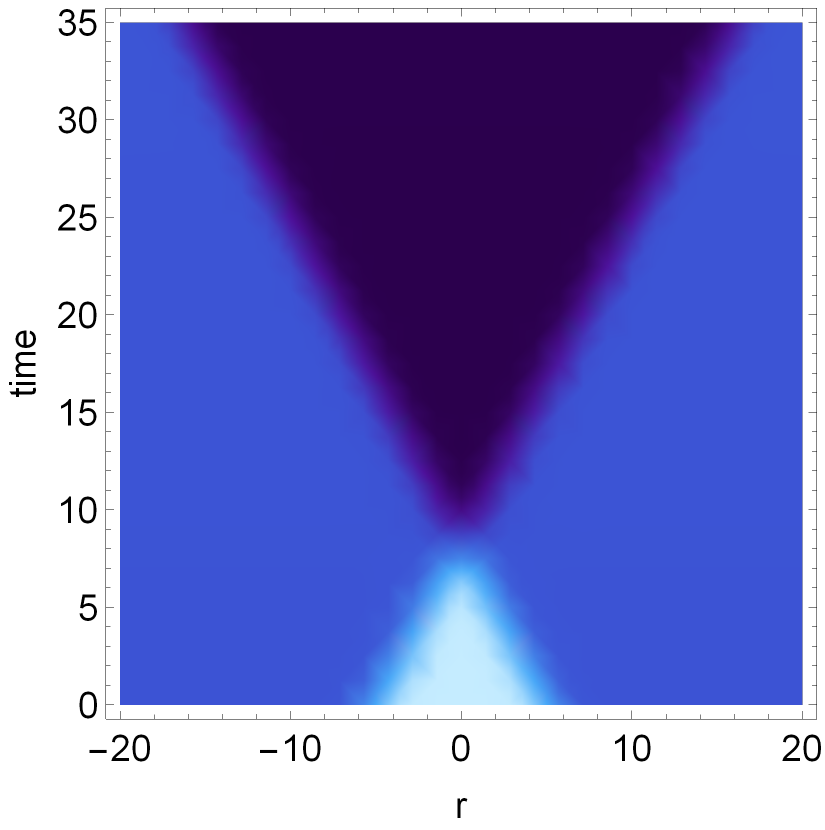}\\
    (c) $v_{in}=0.45$ \hspace{5cm} (d) $v_{in}=0.60$\\
    \includegraphics[height=6cm,width=7.5cm]{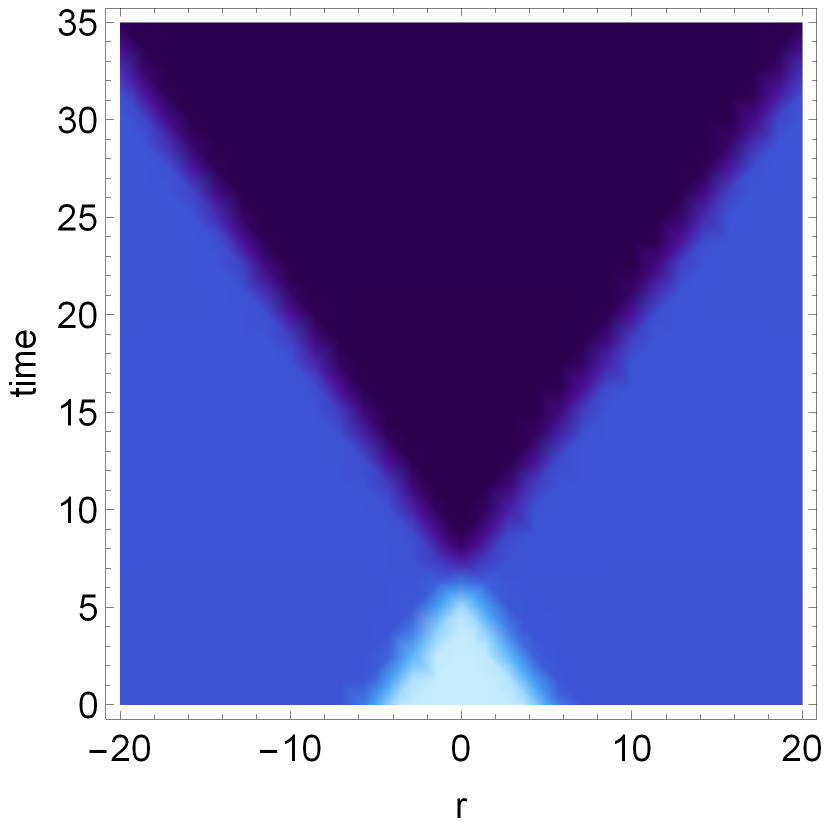}
    \includegraphics[height=6cm,width=7.5cm]{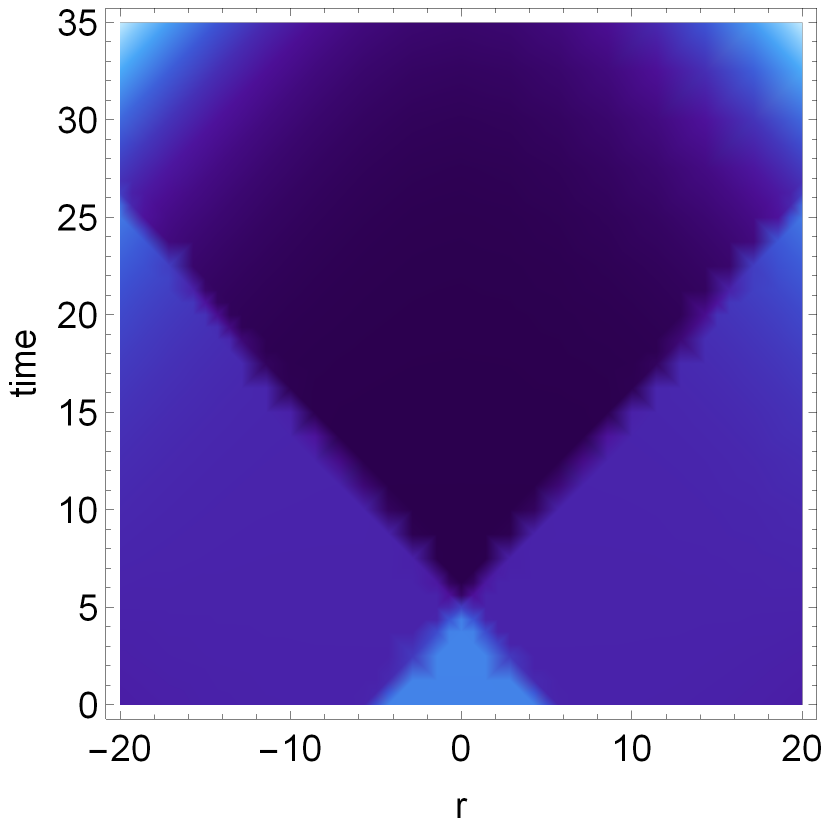}\\
    (e) $v_{in}=0.75$ \hspace{5cm} (f) $v_{in}=0.90$\\
    \vspace{0.1cm}
    \caption{Scattering of the structures.}
    \label{SGfig7}
\end{figure}

\begin{figure}[ht!]
    \centering
    \includegraphics[height=7cm,width=7.3cm]{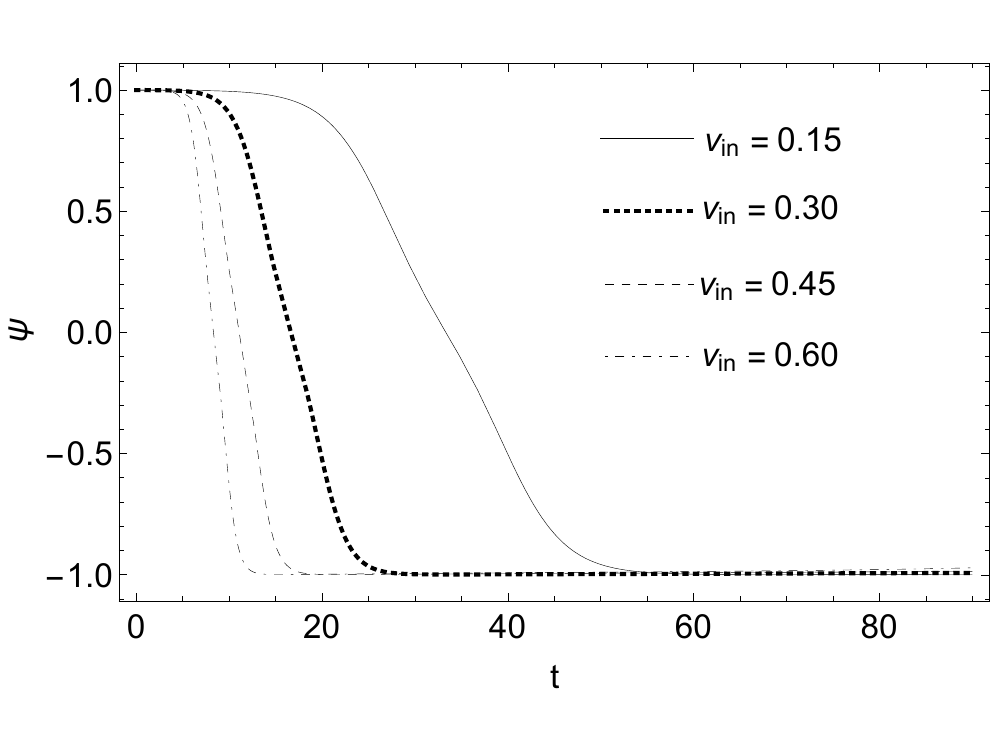}
    \vspace{-0.25cm}
    \caption{Time-dependent scattering of structures from the collision point $r=0$.}
    \label{SGfig8}
\end{figure}

For simulation purposes, one adopts steps of $10^{-4}$ in the numerical computation. Analyzing the numerical results, we found the critical values for the initial velocity ($v_{cr}$), i. e., $v_{cr}\simeq 0.965$. The $v_{cr}$ separates the scattering process into two regimes. If $0.965<v_{in}<0.153$, the structures collide, annihilating each other and radiating their energy. Meanwhile, for $v_{in}>0.153$, one notes a structure scattering (see figure \ref{SGfig7}). The scattering process occurs for speed $0.153<v_{in}<0.965$. In this case, after the collision, one perceives that each structure carries an opposite topological charge, and the interforce is attractive when the structures are far apart (see table I). However, when they are closer together, they repel, reflect, and move away, giving rise to the scattering of an elastic collision. Mathematically, this event class is given by
\begin{align}    \psi_{(-1,+1)}\cup\psi_{(+1,-1)}\to\psi_{(+1,-1)}\cup\psi_{(-1,+1)}.
\end{align}
We expose the illustration of this effect in figures (\ref{SGfig7}) and (\ref{SGfig8}).

\section{Phase transitions of the matter field} 

A natural inquiry is: does the matter field assume other field configurations, such as multi-kink solutions? To answer this question, we will use arguments from information theory, namely, configurational entropy (CE). Indeed, we will use the discussions of a variant from the CE, known as differential configurational complexity (DCC) \cite{Gleiser}. Let us start our analysis by pointing out that CE emerged from the theory proposed in 1948 by Claude E. Shannon \cite{Shannon}. The communication theory proposed by Shannon tells us that
\begin{align}
    S=-\sum_{j}\rho_j\text{ln}\rho_j,
\end{align}
where $\rho_j$ represents the probability $j$-th given the corpus of every possible message. Shannon's information arises pursuing to quantify the maximum rate of information transmission between an addresser and an addressee. In other words, Shannon's entropy gives us the best way for information to propagate in a medium. For more details, see Refs. \cite{Gleiser,Shannon,LA}.

Based on Shannon's concept, the CE concept arises. CE appears as an informational complexity measuring applied the localized field configurations. The message contents of the CE are components of the power spectrum. Thus, to accomplish our purpose, let us use the concept of a CE variant, i.e., the Differential Configurational Complexity (DCC), namely,
\begin{align}
    \mathcal{S}=\int \rho({\bf k})\text{ln}\rho({\bf k})\, d{\bf k}.
\end{align}

As mentioned in Ref. \cite{LA}, DCC contemplates measuring the informational complexity of a localized field configuration. For a field configuration with energy $\mathcal{E}(r)$, one describes the two-dimensional wave modes decomposition by Fourier's transform:
\begin{align}\label{Fourier}
    \mathcal{F}({\bf k})=\frac{1}{\sqrt{2\pi}}\int \mathcal{E}(r)\e^{i\textbf{k}\cdot \textbf{r}} d{\bf r}.
\end{align}

A device sensitive to the full-wave spectrum will detect a wave mode within a volume $d$\textbf{k} centered on {\bf k} with the probability proportional to the mode power. The two-dimensional probability is
\begin{align}
    p({\bf k}\vert d{\bf k})\propto \vert\mathcal{F}(\textbf{k})\vert^2\, d\textbf{k}.
\end{align}
This probability allows us to write the modal fraction as
\begin{align}\label{modal}
    f(\textbf{k})=\frac{\vert\hat{\mathcal{F}}({\bf k})\vert}{\vert\hat{\mathcal{F}}({\bf k_*})\vert},
\end{align}
so that the Differential Configurational Complexity (DCC) \cite{GS} is
\begin{align}\label{DCC}
    \mathcal{S}_{\mathcal{C}}=-\int\, f(\textbf{k})\text{ln}[f(\textbf{k})]\, d\textbf{k}.
\end{align}
The integrand of Eq. (\ref{DCC}) is called entropic density. 

\begin{figure}[ht!]
    \centering
    \includegraphics[height=6cm,width=7.5cm]{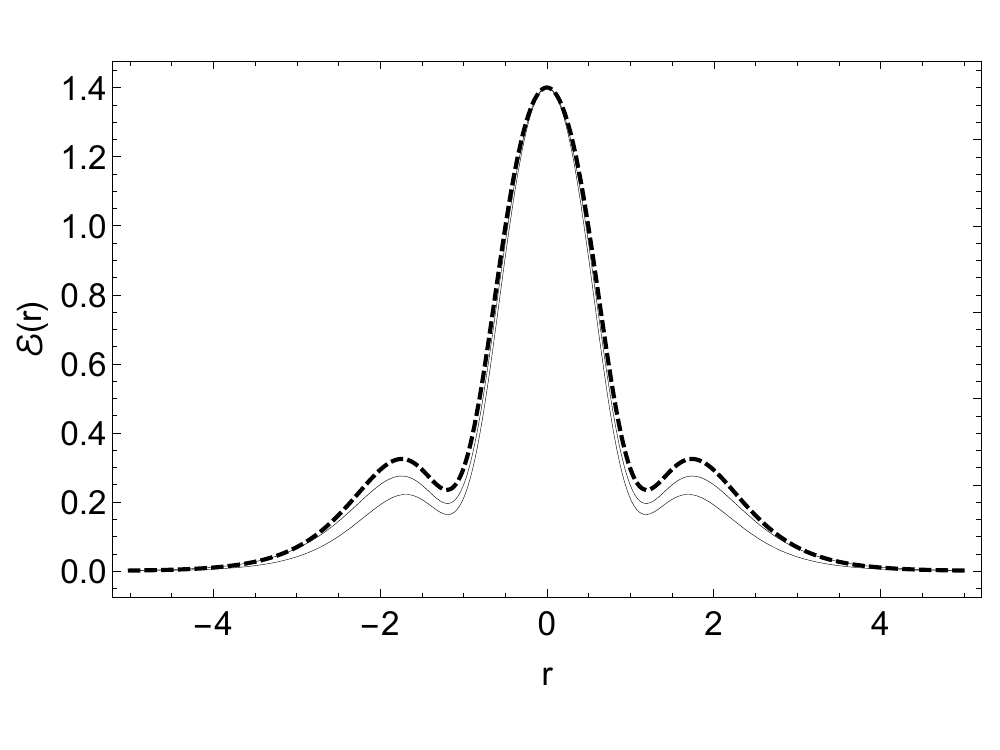}
    \vspace{-0.3cm}
    \caption{Energy density of matter field when $p=1.00$ (dashed line), $1.05$, and $1.10$.}
    \label{fig9}
\end{figure}

It is necessary to investigate the energy density of the matter field to calculate the DCC. In this case, the energy density is 
\begin{align}\label{energy}
    \mathcal{E}(r)=\cosh^{-2p}(r)\bigg[\cosh^{-2p}(r)\psi'(r)+\frac{\lambda}{2}(\nu^2-\psi(r)^2)^2\bigg].
\end{align}
The matter field has the localized energy profile shown in Fig. \ref{fig9}.

To achieve our aspiration of analyzing the phase transitions, let us assume the numerical solutions of the matter field (Fig. \ref{fig2}). With numerical result (\ref{energy}), we calculate the Fourier transform (\ref{Fourier}). Applying the modal fraction (\ref{modal}) and the energy density (\ref{energy}), we numerically calculate the entropic density (\ref{DCC}) and the DCC (Tab. \ref{tab2}).
\begin{figure}[ht!]
    \centering
      \includegraphics[height=6cm,width=7.5cm]{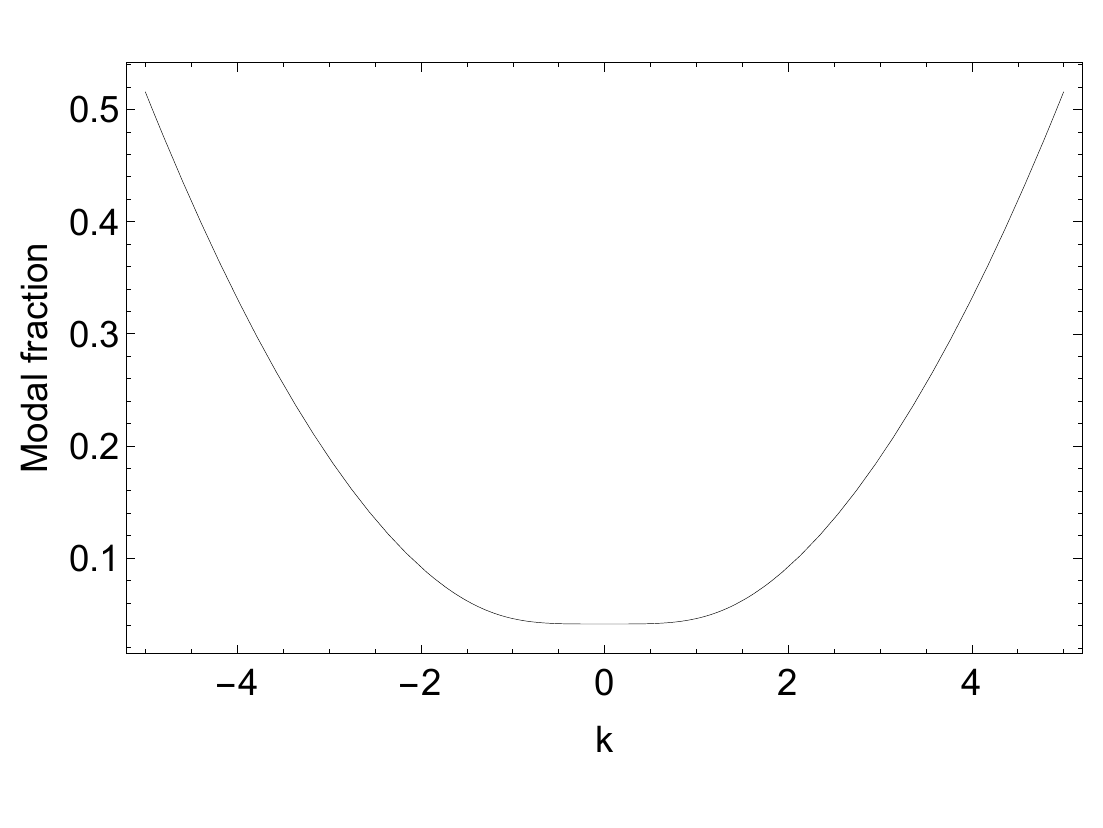}
    \includegraphics[height=6cm,width=7.5cm]{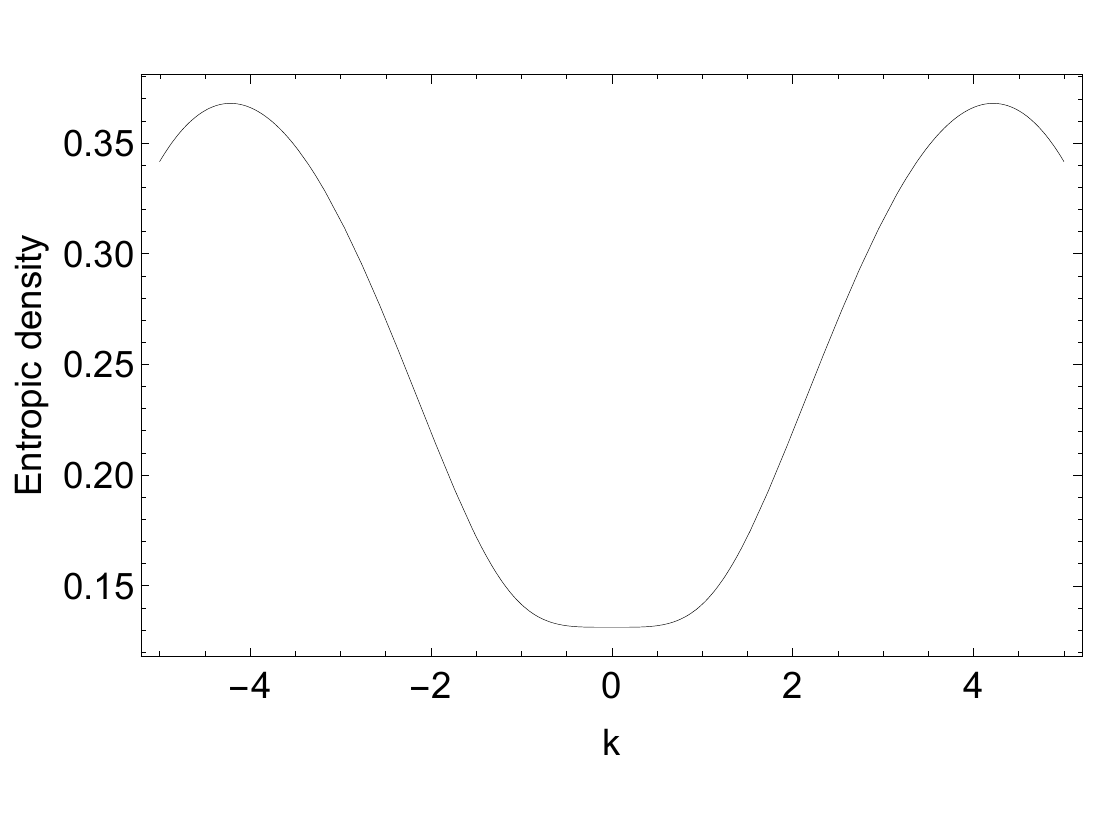}\\
    \vspace{-0.3cm}
    \begin{center}
        (a) \hspace{6cm} (b)
    \end{center}\vspace{-0.3cm}
    \caption{(a) Modal fraction of the structures regardless of the $p$-parameter. (b) Entropic density of the DCC.}
    \label{fig10}
\end{figure}

\begin{table}[ht!]
\centering
\caption{Numerical results of the DCC.}
\label{tab2}
\resizebox{4cm}{2cm}{%
\begin{tabular}{|c|c|}\hline
$p$-parameter & DCC  \\ \hline
  1.00 & 1.83984 \\ \hline
  1.05 & 1.73782 \\ \hline
  1.10 & 1.64806 \\ \hline
\end{tabular}}
\end{table}

The numerical result of the entropic density of the DCC shows an absolute minimum at the origin from the power space, i.e., $k=0$. This minimum occurs in the range of values $\vert k\vert<0.69$. By simulation, it is clear that this point describes an absolute minimum, which indicates the presence of a domain wall around $r=0$. The existence of the domain wall suggests the presence of multiple walls (or at least two) very close together and suggests a double-phase transition. Furthermore, the curved geometry seems to influence the matter field phase transition producing double-kink-like and double-antikink-like solutions shown in Figs. \ref{fig2}(a) and 2(b). Our calculations of the modal fraction and DCC (Fig. \ref{fig10}(a) and \ref{fig10}(b)) guarantee that the matter field should have a profile similar to the class of kinks solutions. Finally, the DCC indicates that the structures are more likely to occur when $p=1$.

\section{Conclusion} 

In this work, we study self-gravitating two-dimensional solutions in dilaton gravity. To investigate the solution stability, we use some arguments by Zhong et al. \cite{Zhong,Zhong1} in dilaton gauge ($\delta\phi=0$). Furthermore, we use a similar approach to reference \cite{Vakhid} to investigate the zero-mode (translational) of the structure. We also calculate the interforce between configurations with opposite topological charges and analyze the scattering phenomenon. The structure phase transitions were analyzed using DCC arguments.

Adopting a canonical matter field in dilaton gravity arises, naturally, the following question: What is the influence of dilaton gravity on the matter field? The answer to this question appears in the results found in the second section. Indeed, the 2D dilaton gravity with the metric (\ref{eq2}) induces deformation in the matter field. In this way, the solutions that describe this field are double-kink-like profiles. This fascinating result is analogous to the results found in the brane-splitting process in cosmological scenarios \cite{LS,Moreira,Ranieri1}. This similarity in the results leads us to believe that the deformation of the kink structures to the double-kink is related to the warp factor profile chosen.

Considering the formation of a pair of structures (widely spaced) with opposite topological charges, we observe that the field configurations suffer the action of an interforce. This interforce is attractive, inducing the scattering process. The structure scattering is particularly interesting because the process will depend on the initial velocity. In our model, the scattering process has two possible results. For initial speed in the range $0.965<v_{in}<0.153$, the structures will collide and annihilate each other, radiating their energy. On the other hand, for initial velocity in the range $0.153<v_{in}<0.965$, we have an elastic scattering process.

The peculiar profile of the matter field appears numerically to resemble a double-kink configuration. Although, indications of the numerical solution and stability are not enough affirmations to guarantee that the topological solutions are true double-kinks. For this analysis, we study the energy profile of the matter field. Furthermore, the arguments from the configurational entropy (precisely, the DCC) help us confirm the previous result. The calculation of energy and DCC suggests that multiple phase transitions (or at least two) occur around the neighborhood of the origin. The domain walls are in a region around the range of values in the power spectrum, namely, $-0.69<k<0.69$. This result allows us to interpret that there is a double-phase transition. DCC data permit us to affirm that our structures are not true double-kink.

It is important to observe that when the p parameter increases, the warp factor e$^{2A}$ becomes more significant. Also, while the p-parameter increases, the field configurations go from a more compact field configuration to a smoother profile. This result implies that the interforce is smaller for $p$ smaller. 

The results obtained throughout this work open new doors for a better understanding of the aspects of the topological structures. That is because they will help us to understand the more complex models in high dimensions. Furthermore, this discussion could be extended to models with two-dimensional modified gravity \cite{Schm}. We hope to bring these discussions in some future works.

\section*{Acknowledgments}

F. C. E. Lima (FCEL) expresses gratitude to the Coordenação de Aperfeiçoamento do Pessoal de Nível Superior (CAPES) for the doctoral scholarship number 88887.372425/2019-00 received from August 2019 to July 2023. Furthermore, F. C. E. Lima acknowledges the Department of Physics from the Universidade Federal do Ceará (UFC) for hospitality. C. A. S. Almeida (CASA) is thankful to the Conselho Nacional de Desenvolvimento Científico e Tecnológico (CNPq), grant number 309553/2021-0, for financial support. Additionally, CASA would like to thank the Print-UFC CAPES program, project number  88887.837980/2023-00, for funding. C. A. S. Almeida acknowledges the Department of Physics and Astronomy at Tufts University for its warm hospitality. The authors thank the anonymous referee for their criticisms, comments, and suggestions.
\section*{Data Availability}
The datasets generated during and/or analysed during the current study are available from the corresponding author on reasonable request.


\end{document}